# Position-based quantum cryptography over untrusted networks

Muhammad Nadeem

Department of Basic Sciences
School of Electrical Engineering and Computer Science
National University of Sciences and Technology (NUST)
H-12 Islamabad, Pakistan

muhammad.nadeem@seecs.edu.pk

**Abstract**
In this paper, we propose quantum position-verification schemes where all the channels are untrusted except the position of the prover and distant reference stations of verifiers. We review and analyze the existing QPV schemes containing some pre-shared data between the prover and verifiers. Most of these schemes are based on non-cryptographic assumptions, that is, quantum/classical channels between the verifiers are secure. It seems impractical in the environment fully controlled by adversaries and would lead to security compromise in practical implementations. However, our proposed formulism for quantum position-verification is more robust, secure and according to the standard assumptions of cryptography. Furthermore, once the position of the prover is verified, our schemes establish secret keys in parallel and can be used for authentication and secret communication between the prover and verifiers.

## 1. Introduction

The central task of position-based cryptography is position-verification. A prover proves to a set of verifiers located at certain distant reference stations that he/she is indeed at a specific position [1]. Unconditional security in classical PBC is impossible because of cloning. The eavesdroppers can copy classical information, manipulate and get desired results before an honest prover. Recently, many authors tried to achieve information-theoretically secure position-based cryptography in quantum settings [2-9]. However, Buhrman *et al* showed that all proposed quantum position-verification schemes are insecure. They proved that position-verification is impossible if the position of the prover is his only credential and he does not have any advantage over eavesdroppers beyond his position while eavesdroppers are allowed to share an arbitrarily large entanglement [8]. They showed that the security of any position-based quantum cryptographic scheme can be destroyed by eavesdroppers through teleporting quantum states back and forth and performing instantaneous nonlocal quantum computation, an idea introduced by Vaidman [10]. However, they proved that if eavesdroppers do not share any entanglement (NO-PE model), then secure PBQC is possible. Furthermore, S. Beigi and R. Konig showed that if eavesdroppers posses an exponential (in n) amount of entanglement then they can successfully attack any PBQC scheme where verifiers share secret n-bit string [11].

In the search of unconditional security, some authors proposed that secure PBQC is possible if the prover and the verifiers pre-share some data [7-9]. However, we will show in section IV that these schemes will remain no more secure if channels between the distant verifiers are insecure. In this paper, we propose that position-based quantum cryptography can be made unconditionally secure even over untrusted networks through entanglement swapping [12]. In our proposed schemes, only position of the prover and the reference stations are secure from adversary while all channels between them are insecure. The only advantage honest prover has over eavesdroppers is publically known pre-shared entangled states with the verifiers. These entangled states can be shared through a source between them that emits labeled pairs of entangled qubits (photons in our case). Furthermore, our schemes require only quantum channels where adversaries can easily be detected by quantum measurement principal and quantum no-cloning theorem [13]. If our QPV schemes are carried out for N times successively and no adversaries are detected, same schemes establish secret keys and can be used for authentication and secret information transfer between the prover and verifiers. Our paper is organized as follows. In section 2, we start with the introduction of quantum position-verification, while the protocol most often used in different QPV schemes, entanglement swapping, is described in section 3. We review existing QPV schemes and analyze them in standard cryptographic settings in section 4, and in section 5 we present our proposed QPV schemes. In section 6 and 7, we extend our QPV schemes to position-based key generation and position-based authentication respectively. Finally, we analyze our schemes under known attacks in section 8 and summarize the paper in section 9.

## 2. Introduction to quantum position-verification

In a general position-verification scheme, an honest prover located at a specified position convinces a set of N verifiers at distant reference stations that he/she is indeed at the specific position. Different verifiers send a secret message and a key to decrypt that message in pieces, that is, each verifier sends a bit of key to P such that all the key bits and the message arrive at the position of P concurrently. If P decrypts the message correctly and sends the result to all verifiers in time, position-verification scheme will enable the verifiers to verify his position jointly. But if one or a set of dishonest provers, not at the specified position, intercept the communication and try to convince verifiers that they are at the specified position, a secure position-verification scheme will enable the verifiers to reject it with high probability. Such secure position-verification is impossible in classical cryptography because of cloning but quantum measurement principle and quantum no-cloning theorem can help in developing secure position-verification schemes. To introduce the idea of quantum position-verification in detail, we will review the basic 1-round QPV scheme $PV_{BB84}$ based on the BB84 encoding [14]. More detailed analysis of this scheme can be found in [8]. Explicit procedure of the scheme for two verifiers follows:

1). $V_0$ prepares two secret random bits $x, y \in [0,1]$ and sends them to $V_1$ through secure channel between them.

2). $V_0$ prepares the qubit $H^y|x\rangle$ and sends it to P. Concurrently, $V_1$ sends the bit $y$ to P such that $H^y|x\rangle$ and $y$ arrive at the same time at P.

3). P measures the qubit in basis $y$ and sends the result to both $V_0$ and $V_1$ immediately.

4). $V_0$ and $V_1$ can verify the position of P by confirming the validity of the result and comparing the arrival time of response.

The authors showed that this scheme is secure only in the No-Pre-shared Entanglement (No-PE) model, where the adversaries do not have pre-shared entangled quantum data but have full power of quantum computing [8]. This scheme can easily be generalized to higher dimensions where multiple verifiers send secret information to P in pieces.

## 3. Entanglement swapping

Entanglement swapping [12] is an interesting extension of teleportation [15], in fact, teleportation of entanglement. It causes two quantum particles to become nonlocally correlated even if they have never interacted. Let Alice posses two particles 1 and 2 and Bob has particle 3 while Charlie keeps particle 4 in his possession. Moreover, suppose Bob and Charlie never met with each other (particles 3 and 4 are initially uncorrelated) but Bob's particle 3 is entangled with Alice's particle 1 while Charlie's particle 4 is entangled with Alice's particle 2 in one of Bell's state:

$$\left|\beta_{u_i u_j}\right\rangle = \frac{|0\rangle|u_j\rangle + (-1)^{u_i}|1\rangle|1 \oplus u_j\rangle}{\sqrt{2}} \quad (3.1)$$

where $u_i, u_j \in [0,1]$ and $\oplus$ denotes addition with mod 2. The initial quantum state of four particles 1, 2, 3 and 4 will be;

$$\left|\beta_{u_1 u_3}\right\rangle \otimes \left|\beta_{u_2 u_4}\right\rangle = \frac{|0\rangle|u_3\rangle + (-1)^{u_1}|1\rangle|1 \oplus u_3\rangle}{\sqrt{2}} \otimes \frac{|0\rangle|u_4\rangle + (-1)^{u_2}|1\rangle|1 \oplus u_4\rangle}{\sqrt{2}} \quad (3.2)$$

By performing Bell state measurement [16] on her particles 1 and 2, Alice can project Bob and Charlie's particles (3 and 4) into one of the four possible Bell states:

$$\left|\beta_{u_1 u_2}\right\rangle \otimes \left|\beta_{u_3 u_4}\right\rangle = \frac{|0\rangle|u_2\rangle + (-1)^{u_1}|1\rangle|1 \oplus u_2\rangle}{\sqrt{2}} \otimes \frac{|0\rangle|u_4\rangle + (-1)^{u_3}|1\rangle|1 \oplus u_4\rangle}{\sqrt{2}} \quad (3.3)$$

Initially, entangled pairs were (1,3) and (2,4). But after BSM by Alice, irrespective of outcome, entangled pairs are (1,2) and (3,4). One can say that particles 3 and 4, initially uncorrelated, become nonlocally correlated through entanglement swapping. In order to complete the protocol, Alice will have to communicate two classical bits (say) to Bob, who can then share a definite bell state $\left|\beta_{u_3 u_4}\right\rangle$ with Charlie after applying suitable unitary local transformations. If initial Bell sates of entangled pair (1,3) and (2,4) are known to Alice, she will be certain about the Bell state of pair (3,4) after performing BSM on qubits 1 and 2. For example, if initial Bell states of entangled pairs (1,3) and (2,4) were $\left|\beta_{01}\right\rangle$ and $\left|\beta_{00}\right\rangle$ and Alice measure particles 1 and 2 in the state $\left|\beta_{10}\right\rangle$, then particles 3 and 4 will be entangled in state $\left|\beta_{11}\right\rangle$. Detailed calculations can be found in appendix. All possible BSM results of Alice and corresponding Bell states of particles 3 and 4 are summarized in table 1. For simplicity, we will write $\left|\beta_{u_i u_j}\right\rangle$ as $u_i u_j$ from now on.

**Table 1:** This table shows all possible initial states of particles 1-4 and corresponding outcomes of BSM on particles 1 and 2. For example, if initial entangled pairs (1,3) and (2,4) were both in states 00 then after BSM on 1 and 2, new entangled pairs (1,2) and (3,4) would be in one of the possible Bell sates: 00 and 00, 01 and 01, 10 and 10, 11 and 11.

| $u_1 u_3 \otimes u_2 u_4$ | | | | $u_1 u_2 \otimes u_3 u_4$ | | | |
|---|---|---|---|---|---|---|---|
| 0000 | 0101 | 1010 | 1111 | 0000 | 0101 | 1010 | 1111 |
| 0001 | 0100 | 1011 | 1110 | 0001 | 0100 | 1011 | 1110 |
| 0010 | 0111 | 1000 | 1101 | 0010 | 0111 | 1000 | 1101 |
| 0011 | 0110 | 1001 | 1100 | 0011 | 0110 | 1001 | 1100 |

**4. Existing QPV schemes containing pre-shared data**

For simplicity, we will discuss all the existing schemes in one dimension. Higher dimensional generalization of these schemes is straightforward and can be found in corresponding references. First we will review these schemes under their proposed assumptions while in our analysis of these schemes; we will consider the standard assumptions of cryptography. That is, eavesdroppers have full control over environment except position of the prover and reference stations. They have unlimited power of receiving, transmitting and manipulating quantum and classical information in no time. Furthermore, they can jam the communication between the honest prover and verifiers.

**4.1. QPV scheme-I**

A. Kent proposed that secure quantum position-verification is possible if the prover and one of the verifiers pre-share some classical bit string unknown to eavesdroppers [7]. This secret data can be then used as a secret key to authenticate the communication. The prover and the verifier can generate longer key $k_0 k_1 k_2 .........$ through quantum key expansion protocol. Moreover, other verifiers still need to communicate some secret information publically with P. The scheme is outlined below:

1). $V_0$ and $V_1$ send randomly chosen bits $x_i$ and $y_i$ from their classical strings $x$ and $y$ respectively. They send this data to P such that these bits arrive at P in pairs, that is, $x_1$ and $y_1$ arrive simultaneously, then $x_2$ and $y_2$, and so on.

2). P retrieves the key bit $k_{4i+2x_i+y_i}$ and sends this bit to both $V_0$ and $V_1$ simultaneously.

3). $V_0$ can verify the position of P if key bit is correct and arrived in time. If P succeeds N times by sending correct bit, $V_0$ authenticates the position of P.

This scheme seems secure but impractical because security of this scheme is based on pre-shared classical secret key which can be expanded through quantum key distribution.

### 4.2. QPV scheme-II

Buhrman *et al* proposed a scheme, $PV_{BB84}^\varepsilon$ EPR version, where one of the verifiers shares an entangled state with the prover [8]. The scheme also requires a secret bit string shared between the verifiers who send this secret information to the prover publically. In one dimension, the scheme is given below:

1). $V_0$ prepares secret random bit $y \in [0,1]$ and sends to $V_1$ through secure channel between them.

2). $V_0$ prepares a two qubit Bell state, keeps one qubit and sends other to P. Simultaneously, $V_1$ sends bit $y$ to P such that both entangled qubit and $y$ reach at P at the same time.

3). P measures the qubit in basis $y$ and sends the result to both $V_0$ and $V_1$ immediately.

4) When measurement result of P arrives, $V_0$ then measures his qubit and sends the result to $V_1$ through secure channel.

5). $V_0$ and $V_1$ can verify the position of P by confirming the validity of the result and comparing the arrival time of response.

Again this scheme is secure only in the No-PE model. In the cryptographic environment where eavesdroppers can possess and share arbitrarily large entangled states, security can be spoofed. Detailed security analysis and higher dimensional version of this scheme can be found in [8].

### 4.3. QPV scheme-III

R. Malaney, proposed a large class of quantum position-verification schemes where different distant verifiers and the prover share entangled data. His work was granted US patent in 2012 [9]. One of his QPV schemes based on entanglement swapping proceeds as below:

1). Let $V_0$ posses an entangled qubit pair (1, 2) and $V_1$ posses an entangled qubit pair (3, 4) in one of the four Bell states, for instance both in 11.

2). At time $t_0$, $V_0$ sends qubit 2 to P and at time $t_1$, $V_1$ sends qubit 3 to P through public channels.

3). P performs a BSM on qubits 2 and 3 and gets one of the Bell states, say 10. This measurement projects the qubits 1 and 4 into Bell state 10, only known to P at the moment. P immediately sends his measurement result to both $V_0$ and $V_1$ simultaneously.

4). Suppose $V_0$ and $V_1$ receive the BSM result from P at time $T_0$ and $T_1$ respectively. $V_1$ immediately transmits his qubit 4, time $T_1$ and BSM result to $V_0$ through secure public channel between them.

5). $V_0$ performs BSM on qubit 1 and 4 and confirms that his result (10) is consistent with that of P.

6). $V_0$ and $V_1$ can verify the position of P if times $T_0$-$t_0$ and $T_1$-$t_1$ are consistent with the position of P.

Unconditional security of this scheme is based on unreal assumption that channels between distant verifiers are secure. If channels between the verifiers are not secure, adversaries can easily break this scheme. They can intercept qubits 2 and 3, process them and can get the secret BSM result from P. Moreover, both $V_0$ and $V_1$ cannot detect the presence of adversaries in this scheme. Cheating scheme by adversaries is shown in figure 1 and is described below:

1). Let $V_0$ possesses an entangled qubit pair (1, 2) and $V_1$ possesses an entangled qubit pair (3, 4) in one of the four Bell states, for instance both in 11. Moreover, suppose eavesdropper $E_0$ lying between $V_0$ and P possesses entangled qubit pair (5, 6) in Bell state 00 while eavesdropper $E_1$ lying between $V_1$ and P have entangled qubit pair (7, 8) also in Bell state 00.

2). At time $t_0$, $V_0$ sends qubit 2 to P but $E_0$ intercepts it and sends her qubit 6 to P. Similarly at time $t_1$, $V_1$ sends qubit 3 to P but $E_1$ intercepts it and sends her qubit 8 to P. Simultaneously, $E_1$ sends qubit 7 to $E_0$.

3). P performs a BSM on qubits 6 and 8 and gets one of the Bell states, say 10. This measurement projects the qubits 5 and 7 into Bell state 10, only known to P at the moment. P immediately sends his measurement result to both $V_0$ and $V_1$ simultaneously.

4). $V_0$ and $V_1$ will receive the BSM result from P at times $T_0$ and $T_1$ as if no adversary is happened. $V_1$ immediately transmits his qubit 4, time $T_1$ and BSM result to $V_0$ but $E_0$ intercepts, as the channel between them is not secure. $E_0$ performs BSM on qubit 5 and 7 and gets 10. Then he applies unitary transformations on qubit 2 such that qubits 1 and 2 get entangled in the state 10, and sends it to $V_0$.

5). $V_0$ will perform BSM on qubit 1 and 2 and confirm that his result is consistent with that of P.

6). Both $V_0$ and $V_1$ will verify the position of P, as if no adversary has happened, as times $T_0$-$t_0$ and $T_1$-$t_1$ are consistent with the position of P.

Since the measurements and timing of eavesdroppers are exactly the same as those of the honest prover, verifiers $V_0$ and $V_1$ cannot differentiate between the honest prover P at a certain position and eavesdroppers at different positions. Hence, eavesdroppers cheat the prover and verifiers without being detected.

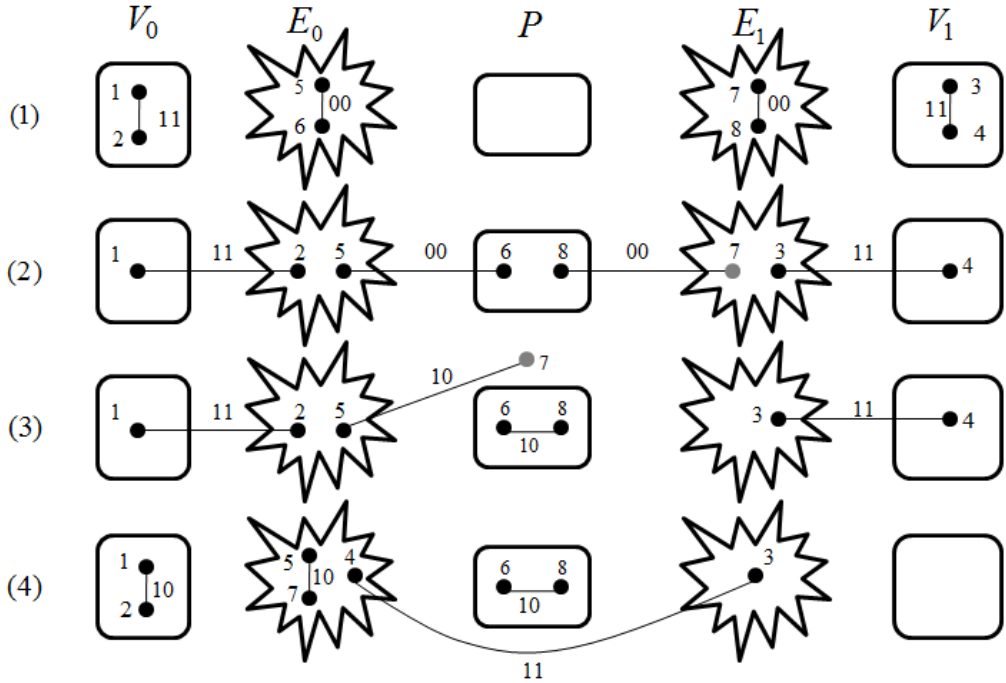

Figure 1: Cheating scheme for QPV scheme-III.

**4.4. QPV scheme-IV**

R. Malani proposed another QPV scheme based on entanglement swapping [9]. This scheme is described as follows:

1). $V_0$ shares two entangled qubit pairs (1, 2) and (3, 4) with the prover P, in one of the four Bell states, for instance both in 00. Suppose he also shares an entangled qubit pair (5, 6) with Bob in the Bell state 11. All this information is public.

2). $V_0$ performs a BSM on qubits 3 and 5 and gets one of the Bell states, say 01. This measurement projects the qubits 4 and 6 into 10, only known to $V_0$.

3). $V_0$ communicates with $V_1$ through a secure public channel between them and informs him about his BSM result, 01. Now $V_1$ also knows his qubit 6 is entangled with P's qubit 4 in the Bell state 10.

4). Both $V_0$ and $V_1$ encode a 2-bit message on their qubits 1 and 6 respectively, through super dense coding [17], and send their encoded qubits to P simultaneously through public channels.

5). P retrieves the encoded 2-bit message by performing BSM on Bell pairs (1, 2) and (4, 6) and immediately sends messages to $V_0$ and $V_1$ through classical channels.

6) $V_0$ and $V_1$ can verify the position of P by comparing the arrival time of response.

Again this scheme assumes that channel between distant verifiers is secure which is not a realistic scenario. In other case, eavesdroppers can intercept and get BSM result of $V_0$, 01. So they will also be able to know that $V_1$ and P have entangled qubit pair in the state 10. Furthermore, eavesdroppers can intercept qubits sent from $V_0$ and $V_1$ and find encoded 2-bit message. The cheating strategy for this scheme is shown in figure 2 and is described below:

1). $V_0$ shares two entangled qubit pairs (1, 2) and (3, 4) with the prover P, in one of the four Bell states, for instance both in 00. Suppose he also shares an entangled qubit pair (5, 6) with Bob in the Bell state 11. Moreover, suppose eavesdropper $E_0$ lying between $V_0$ and P possesses entangled qubit pair (7, 8) in Bell state 00 while eavesdropper $E_1$ lying between $V_1$ and P have entangled qubit pair (9, 10) in Bell state 00.

2). $V_0$ performs a Bell state measurement on qubits 3 and 5 and gets one of the Bell states, say 01. This measurement projects the qubits 4 and 6 into 10, only known to $V_0$.

3). $V_0$ communicates with $V_1$ through an insecure public channel and informs him about his BSM result, 01. Now $V_1$ also knows his qubit 6 is entangled with P's qubit 4 in the state 10. Eavesdroppers intercept and also get this information.

4). Let $V_0$ encodes a 2-bit message 10 on his qubit 1 and $V_1$ encodes a 2-bit message 11 on his qubit 6 respectively, through super dense coding, and send their encoded qubits to P simultaneously through public channels. $E_0$ and $E_1$ intercept these qubits and send their qubits 7 and 9 respectively to P.

5). P performs BSM on Bell pairs (7, 2) and (9, 4) and immediately sends his BSM results 01 to $V_0$ and 10 to $V_1$ through classical channels. $E_0$ and $E_1$ intercept these results, perform BSM on their retained qubits (both 11 say) and they will know 2-bit secret messages of $V_0$ (10) and $V_1$ (11). While decoded message by P will be wrong for sure. Eavesdroppers can jam the signals of p and send exact 2-bit messages to $V_0$ and $V_1$.

6) $V_0$ and $V_1$ will verify the position of P, as if no adversary has happened, by comparing the arrival time of response.

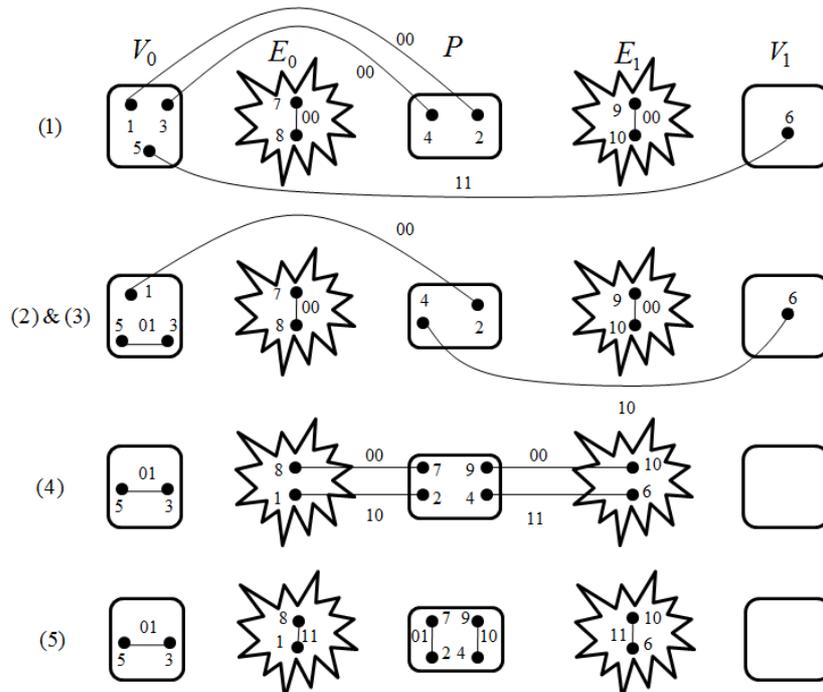

Figure 2: Cheating scheme for QPV scheme-IV.

Thus measurement results and times of $V_0$ and $V_1$ are consistent as if no adversary has happened. Moreover, both $V_0$ and $V_1$ cannot detect the presence of adversaries in this scheme. Hence, eavesdroppers cheat the prover and verifiers without being detected.

**5. Our quantum position-verification schemes**
In this work, we propose quantum position-verification schemes under the more realistic and cryptographically standard assumptions. We assume that the position of the honest prover and reference stations are secure from adversary; enabling them to store and hide the quantum data and process. We also assume that the reference stations are trusted and known to each other. However, quantum/classical channels are not secure; neither between the prover and verifiers nor between different verifiers. Moreover, there is no bound on storage, computing, receiving and transmitting powers of eavesdroppers. In short, eavesdroppers have full control of environment except prover's position and reference stations. We also assume that all reference stations and the prover has fixed position in Minkowski space-time where all verifiers have précised and synchronized clocks. Finally, we suppose that signals can be sent between prover and reference stations at the speed of light. While the time for information processing at position of the honest prover and reference stations is negligible. For simplicity, we will discuss our schemes for one honest prover P and two verifiers $V_0$ and $V_1$ at distant reference stations $R_0$ and $R_1$ such that the prover is at a distance d from both reference stations.

**5.1. QPV scheme-A**
This scheme is shown in figure 3 and its explicit procedure follows:
1). $V_0$ shares two entangled qubit pairs (2, 5) and (3, 7) with the prover P, in one of the four Bell states, for instance both in 01. Let she also shares two entangled qubit pairs (1, 9) and (4,12) with Bob in the Bell state 11. $V_1$ also shares two entangled qubit pairs (6, 10) and (8, 11) with the prover P, in one of the four Bell states, for instance both in 01. All this information is public.

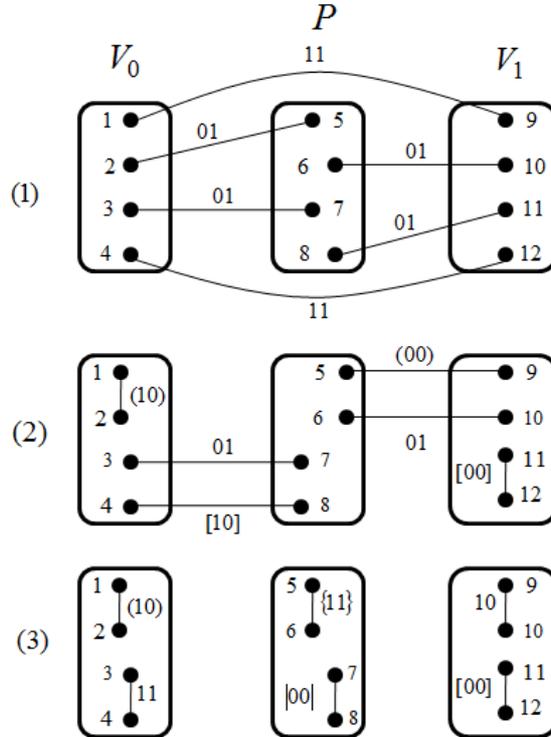

Figure 3: Bell states written as $u_i u_j$ are public. The states $(u_i u_j)$ are known to $V_0$ only while $[u_i u_j]$ are known to $V_1$ only. $\{u_i u_j\}$ are known to both P and $V_0$ while $|u_i u_j|$ are known to P and $V_1$.

2). $V_0$ performs a Bell state measurement on qubits 1 and 2 and gets one of the Bell states, say 10. This measurement projects the qubits 5 and 9 into Bell state 00, only known to $V_0$. Similarly $V_1$ also performs a BSM on qubits 11 and 12 and gets one of the Bell states, say 00. This measurement projects the qubits 4 and 8 into Bell state 10, only known to $V_1$.

3). $V_0$ performs BSM on 3 and 4 and announces result publically, say 11. At this point only $V_1$ knows that the BSM result of P on 7 and 8 will be 00. Concurrently, $V_1$ performs BSM on 9 and 10 and announces result publically, say 10. At this point only $V_0$ knows that the BSM result of P on 5 and 6 will be 11.

4). At time t=0, $V_0$ sends an encoded message to P such that this message can only be decoded with secret 2-bits 11, only known to $V_0$ and P. simultaneously $V_1$ sends an encoded message to P such that this message can only be decoded with secret 2-bits 00, only known to $V_1$ and P.

5). P retrieves the encoded message with corresponding secret 2-bits, obtained by performing BSM on Bell pairs (5, 6) and (7, 8). He immediately sends messages to $V_0$ and $V_1$.

6) $V_0$ and $V_1$ can verify the position of P by comparing the arrival time of response, t = 2d/c.

If verifiers verify the position of P by performing this scheme N time successively, P is identified and his position is authenticated. In this scheme, no secret information is sent publically without properly encoding. The encoded message can only be decoded by P having secret 2-bits.

**5.2. QPV scheme-B**

This scheme is shown in figure 4 and follows:

1). $V_0$ possesses an entangled qubit pair (1, 2) in Bell state 11 and also shares an entangled qubit pair (3, 4), in Bell state 01, with the prover P. $V_1$ also possesses an entangled qubit pair (11,12) in Bell state 11 and shares an entangled qubit pair (9, 10), in Bell state 01, with the prover P. The prover P possesses two entangled qubit pairs (5,6) and (7,8) both in the bell state 00, say. All this information is public.

2). $V_0$, P and $V_1$ perform simultaneously BSM as follows: $V_0$ on qubits 2 and 3, P on qubits 4 and 6, and 8 and 9 while $V_1$ on 10 and 12 respectively. Their BSM results will be known only to them at this stage, for example, 01 to $V_0$, 11 and 01 to P and 10 to $V_1$. Moreover, these measurements will project the qubits 1 and 5 into 00, and 7 and 11 into 01 as shown in figure below. These results will be unknown to everyone.

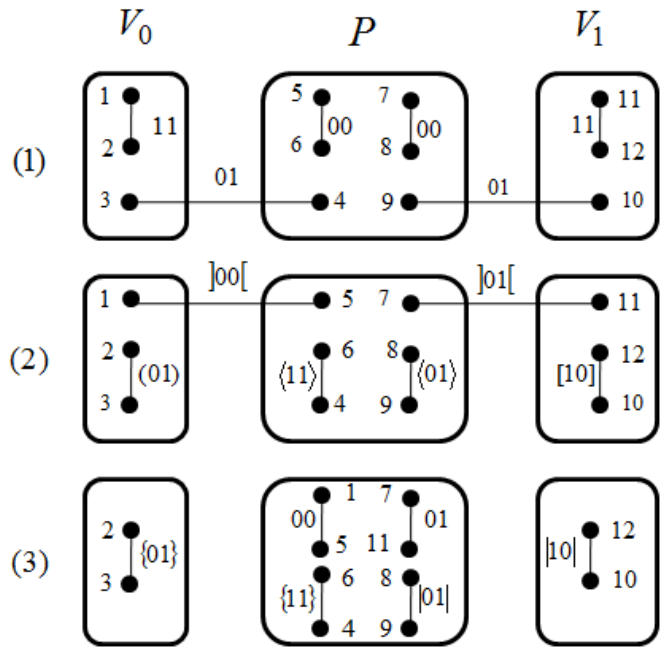

Figure 4: Bell states written as $u_i u_j$ are public. The states $(u_i u_j)$ are known to $V_0$ only, $[u_i u_j]$ are known to $V_1$ only while $<u_i u_j>$ are known to P only. $\{u_i u_j\}$ are known to both P and $V_0$ and $|u_i u_j|$ are known to P and $V_1$. While $]u_i u_j[$ are unknown to everyone.

3). $V_0$ and V1 send their qubits 1 and 11 to P simultaneously. P performs BSM on pairs (1,5) and (7,11) and immediately sends corresponding BSM results 00 and 01 to $V_0$ and $V_1$ respectively.

4). $V_0$ and $V_1$ note round trip time and now they will be aware of corresponding P's BSM results 11 and 01 respectively. Similarly P will be aware of corresponding BSM results of $V_0$ and $V_1$, that is, 01 and 10.

5) $V_0$ sends an encoded message to P such that this message can only be decoded with secret 2-bits 11, only known to $V_0$ and P. simultaneously $V_1$ sends an encoded message to P such that this message can only be decoded with secret 2-bits 01, only known to $V_1$ and P.

6). Only P can retrieve the encoded message by corresponding secret 2-bits. He again encode the same message such that only $V_0$ and $V_1$ can decode it with their secret 2-bits and immediately sends messages to $V_0$ and $V_1$.

7) $V_0$ and $V_1$ can verify the position of P by validating messages and comparing round trip time, $t = 2d/c$.

By using single QPV scheme, verifiers can verify position of p twice; in step 4 and 7. On either stage, if they get wrong response from P, they can detect eavesdroppers in the middle. If verifiers verify the position of P by performing this scheme N time successively, P is identified and his position is authenticated. In both QPV schemes A and B, secret messages can be encoded on qubits by applying arbitrary rotations.

**6. Key establishment in PBQC**

If position of the prover P is verified N times successively, each verifier will have established two different secret keys of length 2N with P, that is,

$$K_V = \{v_1, v_2, ............. v_{2N}\} \quad (6.1)$$

and

$$K_P = \{p_1, p_2, ............. p_{2N}\} \quad (6.2)$$

where $v_i, p_i \in [0,1]$. In our QPV scheme-A, both of the keys $K_V$ and $K_P$ will be known to corresponding verifiers but the prover will know only one of these, $K_P$. However, in our QPV scheme-B, both of the keys $K_V$ and $K_P$ will be known to the prover and corresponding verifiers. These keys can be used further for identification of the prover and authentication of the message transferred. Our position-based key establishment is similar to the one proposed by A. K. Ekert [18] based on shared entanglement states but different in the sense that verifiers and the prover perform BSM on entangled pairs instead of measurement on single entangled particle. A. Cabello also proposed quantum key distribution scheme based on entanglement swapping where distant parties transfer entangled particles through public channels instead of pre-shared entangled states [19]. Eavesdropping attack on Cabello's scheme and further modifications to attain security can be found in following references [20,21].

**7. Authentication in PBQC**

Authentication is a procedure to verify that received message come from the valid entity and has not been altered. Generally authentication can be achieved through following three mechanisms: message encryption (symmetric or asymmetric), message Authentication Code (MAC), or hash functions. Buhrman *et al* introduced the idea of position-based authentication; a message authentication code based on their position-verification scheme [8]. We will show that our proposed QPV schemes can be used as message encryption authentication schemes straightforward.

In the following position-based authentication, we will use photon as a qubit. Horizontally polarized state of photon will be denoted by $|0\rangle$ while vertically polarized state by $|1\rangle$. The scheme works as follows:

1). The prover P chooses a large positive integer *z*, prepares a 2N-qubit state $|\psi\rangle = |0\rangle^{\otimes 2N}$ and generates a classical 2N-bit string $S = \{s_1, s_2, .......s_{2N}\}$ where $s_i$ is any random integer. P encodes string *S* on 2N qubits and sends the state $|\psi_S\rangle$ to $V_0$:

$$|\psi_S\rangle = \otimes_{i=1}^{2N} R(s_i \theta)|0\rangle^{\otimes 2N} \quad (7.1)$$

where $R(s_i\theta)$ is the rotation operator with $\theta = \pi/4^z$.

2) Verifier $V_0$ chooses a different large positive integer $z$ and generates a classical 2N-bit string $T = \{t_1, t_2,....t_{2N}\}$ where $t_i$ is any random integer. $V_0$ encodes string $T$ on $|\psi_S\rangle$ and sends the state $|\psi_{ST}\rangle$ to P:

$$|\psi_{ST}\rangle = \otimes_{i=1}^{2N} R(t_i\theta)R(s_i\theta)|0\rangle^{\otimes 2N} \qquad (7.2)$$

where $R(t_i\theta)$ is the rotation operator with $\theta = \pi/4^z$.

3) P applies rotation $R(-s_i\theta)$ on the state $|\psi_{ST}\rangle$ and then encrypts message $M = \{m_1, m_2,.......m_{2N}\}$; $m_i \in [0,1]$ with his 2N-bit secret key $K_P$ by applying a further rotation $R[(p_i \oplus m_i)\pi/2]$ on $i$th qubit, that is:

$$|\psi_{TM}\rangle = \otimes_{i=1}^{2N} R[(p_i \oplus m_i)\pi/2]R(t_i\theta)|0\rangle^{\otimes 2N} \qquad (7.3)$$

and sends the state $|\psi_{TM}\rangle$ back to $V_0$.

4). To identify the prover P, $V_0$ applies $R(-t_i\theta)$ on the $i$th qubit and measures the state

$$|\psi_M\rangle = \otimes_{i=1}^{2N} R[(p_i \oplus m_i)\pi/2]|0\rangle^{\otimes 2N} \qquad (7.4)$$

in $[|0\rangle, |1\rangle]$ basis. He will get $p_i \oplus m_i$ where $m_i$ can only be retrieved by exact key $p_i$.

5). $V_0$ execute XOR of $p_i \oplus m_i$ and $p_i$. He will get message $M = \{m_1, m_2,.......m_{2N}\}$.

Simultaneously, all other verifiers can perform same scheme with P. Furthermore, all verifiers can note the round trip time of response from P.

## 8. Security analysis

Security of our scheme relies on the fact that no secret information, which could help in spoofing, is sent directly through public channels but is encrypted properly such that only prover and verifiers can decrypt it. In short, proposed QPV schemes remain secure in general and under known entanglement base attacks in particular even if eavesdroppers have infinite amount of pre-shared entanglement and power of non-local quantum measurements in negligible time.

In our QPV scheme-A, eavesdroppers cannot obtain any information about the secret measurement results of $V_0$, $V_1$ and P through public announcements of $V_0$ and $V_1$. Furthermore, eavesdroppers cannot perform intercept/resend or teleportation based attacks because no entangled qubit is transferred between the prover and verifiers. Hence, BSM results are known only to the honest prover and verifiers and only honest prover can respond to verifiers accurately. The verifiers can easily detect adversaries if they try to intercept encrypted communication.

Again in our QPV scheme-B, eavesdroppers cannot get any information about secret BSM results of verifiers and the prover through public announcements of P or by intercepting qubits 1 and 11 sent by $V_0$ and $V_1$ to P over public channels. Suppose eavesdropper between $V_0$ and P possesses already entangled qubit pair (13,14), intercepts qubit 1, performs BSM on 1 and 13 and sends qubit 14 to P. In that case, $V_0$ can easily detect eavesdropper because announcements of P will not be consistent with the BSM results of $V_0$ and P. When $V_0$ will send encoded messages to P in step (5) of the scheme, surely P will decode these messages incorrectly. Similarly P and $V_1$ can detect eavesdropper $E_1$ lying between them.

Finally, our position-based authentication scheme can be made secure by choosing arbitrarily large integer $z$. If $z \gg 1$ (or $\theta \ll 1$), number of non-orthogonal states increases and it becomes impossible to differentiate them, that is, distance between nearest neighbors $\sqrt{1 - |\langle\psi_s(\theta)|\psi_{s+1}(\theta)\rangle|^2}$ approaches to zero. Moreover, only one bit of classical information can be obtained from single qubit [22] while 2N bits are required to identify any randomly chosen $s_i$ (or $t_i$) from 2N-bit string $S$ (or $T$). Hence, the encoding applied in step 1 and 2 acts as a quantum one way function provided $z \gg 1$, only authorized users can extract secret information. For detailed discussion of quantum one way function see [23,24]. Hence, position-based authentication presented in this paper is secure against known attacks; intercept/resend attack, chosen plaintext attack, forward search attack and chosen ciphertext attack.

**9. Conclusion**
In this paper, we review already proposed quantum position-verification schemes based on pre-shared data between the prover and verifiers. QPV scheme-I proposed by Kent seems secure but requires pre-shared classical secret key between the prover and one of verifiers. Scheme-II proposed by Buhrman *et al* is based on pre-shared entangled states between the prover and verifiers but the authors showed that this scheme is secure only if eavesdroppers do not have any entangled data. While we have shown that schemes-III and IV proposed by R. Malani, also based on pre-shared entangled states between the prover and verifiers, are insecure if channels between distant verifiers are not secure.

We proposed two different quantum position-verification schemes to show information-theoretic position-based quantum cryptography is possible even over untrusted networks if the honest prover pre-shares some entangled states with verifiers. Our schemes have numerous advantages over previously proposed schemes in this field. (1). Our proposed schemes are secure even over untrusted networks while all previous schemes may be secure only if channels between distant verifiers are secure. (2). Our schemes verify the position as well as serves as a protocol for position-based QKD which can be used for authentication and communication between the prover and verifiers. However, previously proposed schemes cannot be used for secret communication. For example, in scheme IV, adversaries can spoof position verification as well as get the secret 2-bits of the verifiers. These bits cannot be reused for further communication. (3) Furthermore, in existing scheme IV, verifiers use also classical channels to communicate secret information with the prover in case of N shared entangled pair between them [9]. In principle, eavesdroppers can always monitor classical channels without being detected by authorized users. However, our schemes require only quantum channels while sending secret information. (4) Finally, our proposed QPV schemes can easily detect adversaries while previously proposed schemes can be spoofed by eavesdroppers without being detected.

We presented a formulism that verifies position, establishes secret keys and authenticates the honest prover using a single scheme in position-based quantum cryptography. Our proposed position-based authentication scheme can be modified to more robust authentication mechanism based on hash functions by using secret keys established in our quantum position-verification schemes.

**Acknowledgements**
We are thankful to R. Malaney for introducing us with his work [9].

**Appendix**
In section 3, we write four Bell sates compactly as

$$\left|\beta_{u_i u_j}\right\rangle = \frac{|0\rangle|u_j\rangle + (-1)^{u_i}|1\rangle|1 \oplus u_j\rangle}{\sqrt{2}} \tag{A.1}$$

Where $u_i, u_j \in [0,1]$ and $\oplus$ denotes addition with mod 2. Corresponding four Bell states in $|0\rangle$ and $|1\rangle$ representation will be

$$|\beta_{00}\rangle = \frac{|0\rangle|0\rangle + |1\rangle|1\rangle}{\sqrt{2}} \tag{A.2}$$

$$|\beta_{01}\rangle = \frac{|0\rangle|1\rangle + |1\rangle|0\rangle}{\sqrt{2}} \tag{A.3}$$

$$|\beta_{10}\rangle = \frac{|0\rangle|0\rangle - |1\rangle|1\rangle}{\sqrt{2}} \tag{A.4}$$

$$|\beta_{11}\rangle = \frac{|0\rangle|1\rangle - |1\rangle|0\rangle}{\sqrt{2}} \qquad \text{A.5}$$

Suppose qubit pairs (1,3) and (2,4) are entangled in Bell sates $|\beta_{01}\rangle_{13}$ and $|\beta_{00}\rangle_{24}$. Initially, these four particles will be in state

$$|\beta_{01}\rangle_{13} \otimes |\beta_{00}\rangle_{24} = \frac{|01\rangle_{13} + |10\rangle_{13}}{\sqrt{2}} \otimes \frac{|00\rangle_{24} + |11\rangle_{24}}{\sqrt{2}} \qquad \text{A.6}$$

$$|\beta_{01}\rangle_{13} \otimes |\beta_{00}\rangle_{24} = \frac{1}{2}\left(|01\rangle_{13}|00\rangle_{24} + |01\rangle_{13}|11\rangle_{24} + |10\rangle_{13}|00\rangle_{24} + |10\rangle_{13}|11\rangle_{24}\right) \qquad \text{A.7}$$

$$|\beta_{01}\rangle_{13} \otimes |\beta_{00}\rangle_{24} = \frac{1}{2}\left(|00\rangle_{12}|10\rangle_{34} + |01\rangle_{12}|11\rangle_{34} + |10\rangle_{12}|00\rangle_{34} + |11\rangle_{12}|01\rangle_{34}\right) \qquad \text{A.8}$$

By simple algebraic tricks (adding and subtracting terms like $|u_i u_j\rangle_{12}|u_i u_j\rangle_{34}$), we will get

$$|\beta_{01}\rangle_{13} \otimes |\beta_{00}\rangle_{24} = \frac{1}{2}\left(|\beta_{00}\rangle_{12}|\beta_{01}\rangle_{34} + |\beta_{01}\rangle_{12}|\beta_{00}\rangle_{34} - |\beta_{10}\rangle_{12}|\beta_{11}\rangle_{34} - |\beta_{11}\rangle_{12}|\beta_{10}\rangle_{34}\right) \qquad \text{A.9}$$

By performing Bell state measurement on particles 1 and 2, Alice can project particles 3 and 4 into one of the four possible Bell states; $|\beta_{00}\rangle$, $|\beta_{01}\rangle$, $|\beta_{10}\rangle$ or $|\beta_{11}\rangle$.